\documentclass[usenatbib]{aa}
\usepackage{float,txfonts}
\usepackage[colorlinks]{hyperref}
\begin{document}

\newcommand{\gaea}{{\sc gaea}}
\newcommand{\pmill}{{\sc planck millennium}}
\newcommand{\pms}{PMS}
\newcommand{\hmsun}{\, h^{-1}{\rm M}_\odot}
\newcommand{\msun}{\, {\rm M}_\odot}
\newcommand{\msunyr}{\, {\rm M}_\odot\ {\rm yr}^{-1}}
\def\lesssim{\lower.5ex\hbox{$\; \buildrel < \over \sim \;$}}
\def\gtrsim{\lower.5ex\hbox{$\; \buildrel > \over \sim \;$}}

\title{Do z>6 quasars reside in protoclusters?}
\author{Fabio Fontanot\inst{1,2}\fnmsep\thanks{e-mail:fabio.fontanot@inaf.it}
  \and Roberto Decarli\inst{3}
  \and Gabriella De Lucia\inst{1,2}
  \and Olga Cucciati\inst{3}
  \and Lizhi Xie\inst{4}
  \and Michaela Hirschmann\inst{5,1}
}
\institute{INAF - Astronomical Observatory of Trieste, via G.B. Tiepolo 11, I-34143 Trieste, Italy\label{inst1}
  \and IFPU - Institute for Fundamental Physics of the Universe, via Beirut 2, 34151, Trieste, Italy\label{inst2}
  \and INAF - Osservatorio di Astrofisica e Scienza dello Spazio di Bologna, Via Piero Gobetti 93/3, 40129 Bologna, Italy\label{inst3}
  \and Tianjin Astrophysics Center, Tianjin Normal University, Binshuixidao 393, 300384, Tianjin, China\label{inst4}
  \and EPFL - Institute for Physics, Laboratory for Galaxy Evolution, Observatoire de Sauverny, Chemin Pegasi 51, 1290 Versoix, Switzerland\label{inst5}
}

   \date{Received ???, 2024; accepted ???, 2024}
   
   \abstract{We discuss the properties of a sample of z>6 bright
     (bolometric luminosity L$_{\rm bolo}$>10$^{46.25}$ erg/s) Quasars
     drawn from a realisation of the GAlaxy Evolution and Assembly
     (\gaea) model coupled with the Planck Millennium Simulation. We
     focus on the properties and environment of host galaxies on a
     physical scale of 10 cMpc, and their evolution down to z=0, with
     the aim of assessing how well the bright high redshift QSOs
     population traces the progenitors of the most massive haloes in
     the local Universe. {\gaea} predicts that z>6 bright QSOs live in
     a variety of environments, and that secular processes like disc
     instability are responsible for triggering roughly the same
     number of QSOs as galaxy mergers. We consider mock fields built
     around these high-z QSOs, and we show that roughly half of them
     include other active galaxies (with L$_{\rm bolo}$>10$^{44}$
     erg/s). The predicted large field-to-field variance in the number
     of companions is qualitatively consistent with recent results
     from JWST observations. Descendants of host galaxies at z=0 cover
     a wide range of physical properties and environments with only a
     small fraction of them belonging to massive galaxy clusters
     (M$_{\rm DM}$>10$^{14.5} \msun$). Viceversa, \gaea~predicts that
     only a small fraction of Bright Central Galaxies have a bright
     z>6 QSOs among their progenitors. Our results suggest that
     luminous high-z QSO loosely trace the progenitors of low-z galaxy
     clusters, but that additional information about their environment
     (like the number of active galaxy companions) are required to
     identify the most promising proto-cluster candidates.}


\titlerunning{High-z QSOs in \gaea}\authorrunning{F.~Fontanot et al.}
   
   \maketitle
%
\section{Introduction}
\label{sec:intro}
Quasars (QSOs) observed in the first Gyr of the Universe ($z\gtrsim
5.7$) are among the most active sources emerging from the dark cosmic
ages. They are characterized by rapid accretion rates ($\dot {\rm
  M}$>10$\,\msunyr$) onto black holes that have already assembled
masses as large as 10$^9\,\msun$, and by intense episodes for star
formation (with rates SFR=100--1000$\,\msunyr$) fueled by copious
reservoirs (10$^{10}\,\msun$) of molecular gas (see \citealt{Fan23}
for a recent review). In order to explain this rapid growth of both
black holes and their host galaxies, a common broad-brush picture
postulates that these sources should reside in some of the most
massive dark matter haloes (DMHs) that have already collapsed at these
epochs (i.e. with masses M$_{\rm DM}$ in excess of a few 10$^{12}
\msun$ -- see e.g. \citealt{WyitheLoeb03}). The rationale is that the
estimated expected abundance of cold gas in these environments can
sustain the efficient growth of super-massive Black Holes (SMBHs)
\citep[see e.g.][]{EfstathiouRees88}.  If this is the case, one should
expect a relevant number of close companion galaxies around these
high-z bright sources \citep{Overzier09}.  Observational support to
this hypothesis has been elusive, due to limitations on sensitivity,
size of the field of view, sample sizes, and contradictory findings
connected with selection effects (see discussion in
\citealt{Mazzucchelli17}).

The first spectroscopic confirmations of galaxies in the close
environment of QSOs at $z\gtrsim 5.7$ came from the Atacama Large
Millimeter Array (\citealt{Decarli17}; see also
\citealt{Trakhtenbrot17}; \citealt{Willott17}, \citealt{Neeleman19})
and from integral field spectroscopy with MUSE on the Very Large
Telescope (\citealt{Farina17}, \citealt{Meyer22}). While instrumental
to demonstrate that (at least some) high-$z$ QSOs do live in rich
environments, these studies could only probe the close, $\lesssim
100$\,kpc, scales. In the last few years, more fields have became
accessible thanks to the superb slit-less spectroscopy capabilities of
JWST. \citet{Kashino23}, \citet{Wang24}, \citet{Champagne25a} and
\citet{Wang26} presented studies of the $\sim$10 Mpc$^2$ area around
QSOs at cosmic dawn. The NIRCam Wide Field Slit-less Spectroscopy
(WFSS) mode delivered spectra of the rest-frame blue optical emission
of all the sources in the targeted QSO fields. Via the bright [O{\sc
    iii}] and H$\beta$ line emission, the identification of the
properties of galaxies around the QSOs is straightforward. These new
campaigns have revealed tens of companion galaxies at separations as
large as a few Mpc from the QSO. Significant field-to-field variation
has been reported \citep{Kashino23}. \citet{Eilers24} used the
statistics of companion galaxies around $z\gtrsim 5.7$ QSOs to
estimate the QSO--galaxy correlation function and infer the mass of
the host DMH. They found a minimum host halo mass of log\,M$_{\rm
  halo}/{\rm M_\odot}=12.43_{-0.15}^{+0.13}$.

On the theoretical side, work based on hydro-simulations \citep[see
  e.g.][]{DiMatteo12, Costa14, Habouzit19} has confirmed the need of
efficient SMBH growth in M$_{\rm DM}$ $\gtrsim $ 10$^{12} \msun$
haloes as the most plausible explanation for the emergence of
z$\gtrsim$6 bright QSOs. Several physical mechanisms have been
proposed to explain the observed variety in the number of detected
companions within a $\sim 1 Mpc$ distance of known z$\gtrsim$6 QSOs
(and in particular the dearth of detected companions in some fields),
such as strong radiative stellar feedback \citep{Costa19}, AGN
feedback \citep{Habouzit19}, baryonic physics \citep{Ren21},
luminosity scatter \citep{Zana22} or observational biases
\citet{Zana23}. On the other hand, using a semi-analytic approach
based on the {\sc galform} model, \citet{Fanidakis13} suggested a
slightly different range of DMH masses for the hosts of high-z bright
QSOs (M$_{\rm DM} \gtrsim$ 10$^{11} \msun$); this evidence would
naturally explain some the observed differences in the number of
companions in QSO fields, as a wider range of host halo masses would
correspond to a larger variance in the expected number of bright
companion galaxies.

Much of this previous work refers to the environmental properties of
high-z QSO host on relatively small physical scales. For example,
\citet{Habouzit19} showed that the number count excess in the
environment of high-z QSOs is significant only on scales closer than a
few Mpc to the reference source. A different, but still related, open
question still holds: do these bright high-z QSOs reside at the core
of some of the most prominent overdensities emerging at these early
cosmic ages, often called ``proto-clusters''?  (see
\citealt{Overzier16} for a review) Are they the ``peak of the
iceberg'', i.e., do they eventually mark the highest density regions
of the large--scale structures hosting the QSOs? If QSOs indeed reside
in prominent structures in the first Gyr since the Big Bang, it is
tempting to postulate that they mark the seeds for massive
proto-clusters before cosmic noon and giant galaxy clusters in the
local Universe (see e.g. \citealt{Stiavelli05, Dannerbauer14,
  Morselli14, Hennawi15, Cucciati18, Shah24}). Furthermore, the
tumultuous growth of black holes and their hosts in the early
Universes is expected to be connected with a rapid depletion of cold
gas that would otherwise be used at later times. Hence, QSO hosts are
often thought to represent the progenitors of massive and passive
galaxies, and perhaps of bright central dominant galaxies. By
analysing the outputs of a large N-body cosmological simulation,
\citet{Angulo12} showed that the growth of a z>5 M$_{\rm DM} \sim$
10$^{12} \msun$ DMH into a z$\sim$0 galaxy cluster is connected to a
larger degree with its environment (as defined on scales larger than
$\sim$ 10 Mpc), than to its actual M$_{\rm DM}$, as it is the former
to determine its overall mass assembly. The connection between bright
high-z QSOs and protoclusters can be quantitatively tested by means of
theoretical models of galaxy evolution. However, such studies require
large cosmological volumes to sample the range of possible large scale
environments and evolutionary paths of such rare objects.

Semi-Analytic models (SAMs -- see \citealt{Baugh06} for a review)
represent an ideal framework for this analysis, as they provide a fast
and computationally efficient approach to galaxy evolution, aimed at
following the main physical properties of galaxy populations over a
wide redshift range. SAMs are theoretical tools aimed at providing
predictions for the growth of galaxy populations across cosmic epochs,
starting from a statistical description of the evolution of DMHs in a
cosmological volume. Galaxy evolution involves the baryonic gas
accumulated into the DMH potential wells, that evolves under the
effect of a complex network of physical mechanisms that drive the
consumption of the cold gas reservoir either by forming stars or by
accretion onto SMBHs and the energy release via feedback effects. The
approach relies on approximated parametrizations (empirical, numerical
or purely theoretical in nature) to model the key physical
processes. The lack of an explicit treatment of the gas dynamics
largely reduces the numerical requirements, thus allowing for a
detailed exploration of the parameter space associated with the
assumed prescriptions. The SAM approach cannot track the spatial
distribution of the different baryonic components (to the level of
detail offered by a hydro-simulation), nonetheless, it is an ideal
tool to explore galaxy evolution over a cosmological volume and across
all cosmic epochs, providing the needed flexibility to study the
impact of individual physical mechanisms. In particular, the SAM
approach allows us to sample cosmological volumes large enough
(i.e. of the order of a Gpc$^3$) to find bright z$\gtrsim$6 QSOs in
sizeable numbers and to follow the evolution of the their host
galaxies descendants down to z-0.

In this paper, we will take advantage of the latest version of the
GAlaxy Evolution and Assembly (\gaea) model \citep{DeLucia24},
featuring an advanced modelling for gas accretion onto SMBHs at the
centre of galaxies. In \citet{Fontanot20}, we tuned the model to
reproduce the space density evolution of the Active Galactic Nuclei
(AGN) population at z$\lesssim$4 in different luminosity ranges (the
so-called AGN-downsizing). \citet{DeLucia24} showed that our modelling
of gas accretion and AGN feedback is instrumental (altogether with the
improved modelling of star formation based on the partition of cold
gas into its atomic and molecular components -- \citealt{Xie17}) to
reproduce the observed fractions of quenched galaxies over a wide
range of cosmic epochs. In \citet{Grazian24}, \gaea~predictions have
been compared with a detailed determination of the observed AGN
luminosity function at z$\sim$5, showing a good agreement with
data. In a recent work, \citet{Cammelli25} coupled a modified version
of the \gaea~model with predictions from fast methods based on the
Lagrangian Perturbation Theory \citep{Monaco02} to study the
properties of the high-z AGN population. While the aim of this work
was to explore the impact of different BH seeding mechanisms on the
galaxy/AGN populations, it has also provided some predictions for the
expected space densities and for the shape of the relation between the
mass of the SMBH (M$_{\rm BH}$) and the stellar mass of the host
galaxy (M$_\star$). The main limitation of the \citet{Cammelli25} work
with respect to the questions addressed here is the limited size of
the simulated box (60 cMpc per size, insufficient for detecting a
significant number of bright high-z QSOs). In this paper, we use the
\pmill~simulation \citep[\pms~hereafter]{Baugh19}, that spans a volume
of $\sim$0.5 Gpc$^3$. The PMS allows us to explore a cosmological
volume large enough to detect a sizeable sample of bright QSOs at
high-z, study the physical properties of their host galaxies and
explore their evolution to lower redshift.

This paper is organized as follows. In Section~\ref{sec:simsam} we
present the \gaea~model run on merger trees extracted from the
\pms~simulation, while in Section~\ref{sec:agnsel} we discuss the
definition of a reference high-z mock sample. In Section~\ref{sec:env}
we present our results on the environment of high-z QSOs in GAEA and
compare with recent observational determinations. Then in
Section~\ref{sec:evo} we study the lower redshift evolution of high-z
QSOs and in Section~\ref{sec:z0} we focus on the properties of their
descendants. Finally, we summarize our conclusions in
Section~\ref{sec:discconcl}.

\section{Galaxy formation model}
\label{sec:simsam}

In this study we consider predictions from the latest rendition of the
GAlaxy Evolution and Assembly (\gaea) model, as introduced by
\citet[see also \citealt{Xie24}]{DeLucia24}, featuring a much needed
synthesis of independent developments. In particular, this version of
the model builds on the original model presented in
\citet{Hirschmann16} and includes (i) the explicit partition of cold
gas into atomic and molecular component presented in \citet{Xie17},
(ii) the modelling of cold gas accretion onto SMBHs and AGN feedback
from \citet{Fontanot20} and (iii) the treatment of dynamical processes
acting on satellite galaxies described in \citet{Xie20}. This new
version of the model has been calibrated against the evolution of the
galaxy stellar mass function up to z$\sim$3; the evolution of the AGN
luminosity function up to z$\sim$4 and the local HI and H$_2$ mass
functions. Our previous work has shown that this realisation
reproduces well the fractions and space densities of quiescent
galaxies up to z$\sim$4 \citep{DeLucia24}, the emission line
properties of different galaxy populations \citep{Scharre24}, the
clustering properties of galaxy populations as a function of stellar
mass, star formation activity, HI content and redshift
\citep{Fontanot25}. In a recent paper, \citet{Cantarella25} tested
\gaea~predictions against the latest JWST measurements of the
Ultra-Violet luminosity functions at z>5 and of the evolution of the
mass-metallicity relation finding reasonable agreement for model
predictions with both observational determinations up to z$\sim$9-10.

In the following we will consider the most recent \gaea~realisation
\citep{Fontanot25}, that couples the model with merger trees extracted
from the \pms. The \pms~tracks the evolution of the Large Scale
Structure in a 800$^3$ Mpc$^3$ volume assuming a cosmological model
($\Omega_\Lambda=0.693$, $\Omega_m=0.307$, $\Omega_b=0.04825$,
$n=0.9611$, $\sigma_8=0.8288$, $H_0=67.77 \, {\rm km/s/Mpc}$) based on
the first year results from the Planck satellite
\citep{Planck_cosmpar}. This numerical simulation has been run using
the N-body code {\sc gadget} \citep{Springel05} and approximately 128
billion particles to recover the assembly of the DMHs, resulting in a
particle mass resolution of $1.06 \times 10^8 \hmsun$; running
\gaea~on the \pms~merger trees thus allow us to resolve the physical
properties of model galaxies more massive than $10^8 \msun$. The
\pms~is beneficial for high-redshift studies, with respect to the
other simulations in the Millennium suite (such as the {\sc
  millennium} and {\sc millennium-II}), thanks to its better mass and
time resolution, as well as its larger volume. For more details on the
simulation setting, we refer the reader to the original paper
\citep{Baugh19}.

Given the focus of this work, we will briefly recap the main features
of the modelling of gas accretion onto SMBHs in \gaea. We refer the
interested reader to \citet{Fontanot20} for a detailed description of
the model and its main predictions. First of all, we assume a flat BH
seeding scheme: each time a new DMH is resolved (at 2 $\times$ 10$^9
\hmsun$, corresponding to 20 DM particles) in the merger tree, we seed
it with a $\sim$10$^2 \msun$ BH. We assume that AGN activity is
triggered by physical processes able to destabilise the cold gas in
star forming galaxy discs, make it lose its angular momentum and
infall towards the centre of the galaxy to form a reservoir around the
SMBH. In \gaea~we assume that either galaxy mergers or secular
processes like disc instabilities are possible triggers of AGN
activity. Whenever one of these processes occurs, we take a fraction
of the cold gas available in the model galaxy and we move it into a
reservoir around the central SMBH (whose mass is $M_{\rm rsv}$). The
growth rate of the cold gas reservoir is assumed to be proportional
either to the SFR triggered by the galaxy merger or to the bulge
growth rate associated to the disk instability event via the parameter
$f_{\rm lowJ}$. The gas accretion from the reservoir onto the SMBH
($\dot{M}_{\rm BH}$) is regulated by a viscous accretion timescale
\citep{Umemura00, Granato04}:

\begin{equation}\label{eq:bhaf06}
  \dot{M}_{\rm BH} = f_{\rm BH} \frac{\sigma_B^3}{G} \left(
  \frac{M_{\rm rsv}}{M_{\rm BH}} \right)^{3/2} \left( 1+\frac{M_{\rm
      BH}}{M_{\rm rsv}} \right)^{1/2}
\end{equation}

\noindent
where $\sigma_B$ represents the velocity dispersion of the bulge and
$f_{\rm BH}$ is a free parameter. We limited accretion to a maximum
value of 10 times the Eddington rate \citep{Takeo19, Inayoshi16,
  Delvecchio20}. Finally, AGN activity is assumed to provide feedback
to the host galaxy, by means of AGN-driven outflows, characterized by
a mass loading factor proportional to the SMBH accretion rate, through
a free parameter $\epsilon_{\rm qw}$, following the scaling relation
proposed by \citet{Fiore17}. The mass loading factor quantifies the
cold gas mass removed from the host galaxies and restored to the hot
gas component of the parent DMHs. The adopted values for the key
parameters ($f_{\rm lowJ}$, $f_{\rm BH}$ and $\epsilon_{\rm qw}$) have
been calibrated by requiring model predictions to reproduce the
redshift evolution of the AGN bolometric luminosity function up to
z$\sim$4 (see Tab.~1 in \citealt{DeLucia24} for a complete list of the
main {\gaea} parameters). Bolometric luminosities for our model AGNs
have been computed from the corresponding accretion rates assuming a
15 per cent radiative efficiency \citep{Shankar19nat}.

\section{Selection of high-z QSOs in GAEA}
\label{sec:agnsel}
\begin{figure}
  \centerline{ \includegraphics[width=8cm,height=12cm]{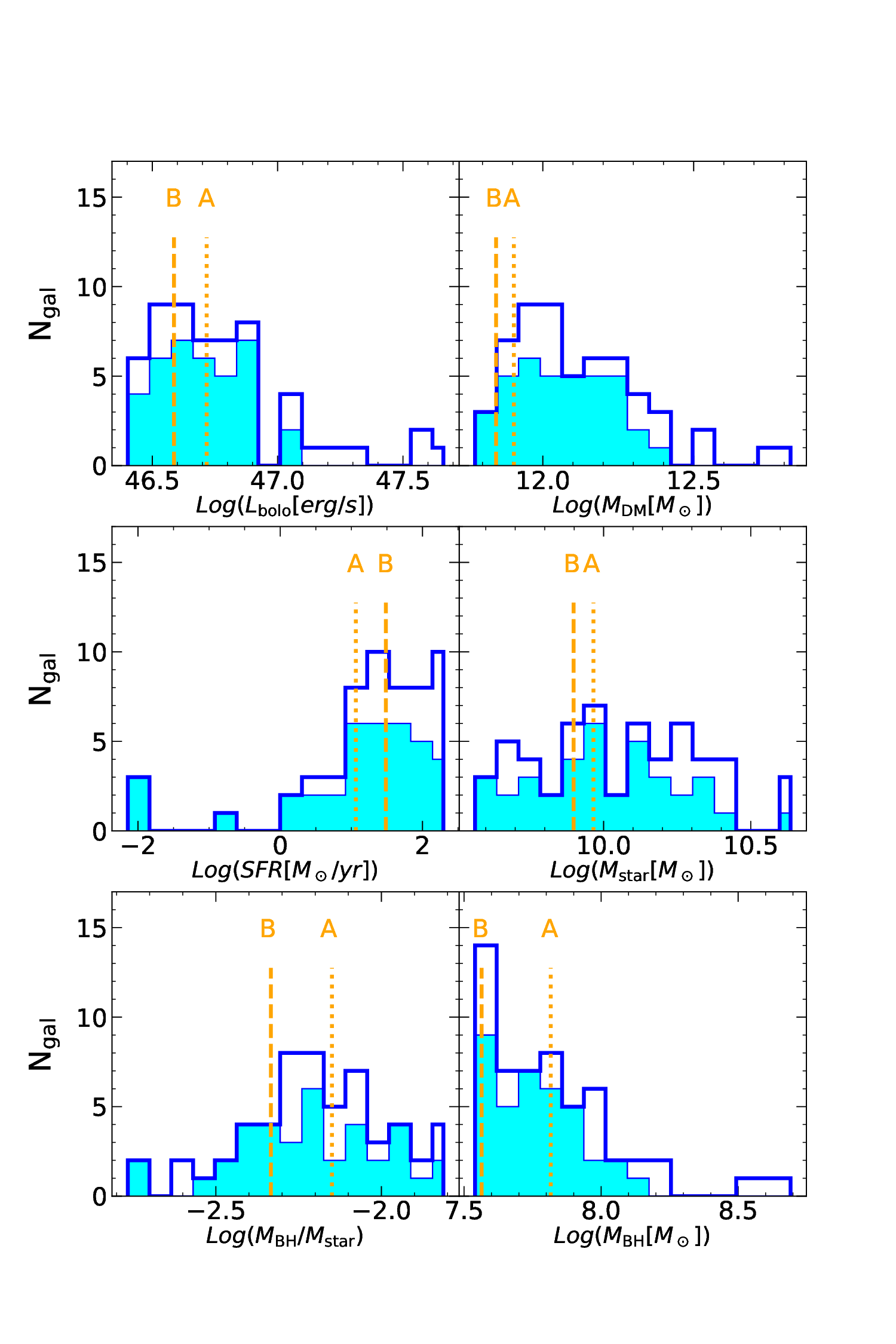}}
  \caption{Properties of the most luminous z>6 QSOs and their host
    galaxies in \gaea. Blue empty histograms refer to the properties
    of the full sample of 56 high-z QSOs, while the cyan filled
    histograms show the same properties for the subsample of QSOs
    triggered by disc instability events. Vertical dashed and dotted
    lines mark the position of our reference objects (see main text
    for more details).}\label{fig:prop}
\end{figure}
\begin{figure*}
  \centerline{ \includegraphics[width=18cm]{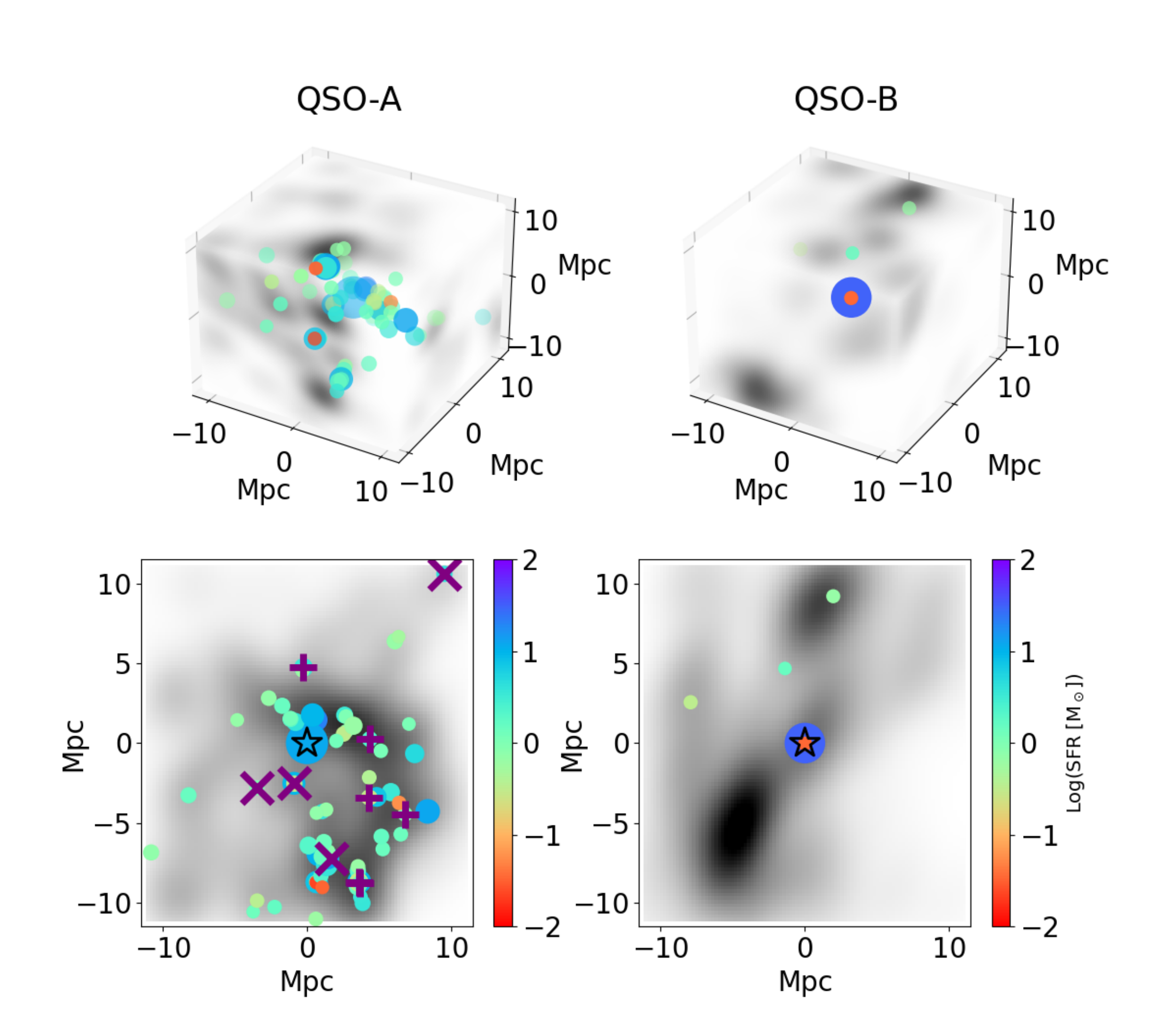}}
  \caption{{\it Upper panels:} 3D galaxy distribution around two
    representative high-z QSOs at z$\sim$7.64. Left-hand panels
    correspond to QSO-A, while right-hand panels to QSO-B (see text
    for more details). Circles mark the position of galaxy companions
    with M$_\star$>10$^{8} \msun$ in a 20 cMpc size box centred on the
    high-z QSO. The size of the symbols is proportional to the galaxy
    stellar mass, and they are colour coded according to their SFR as
    in the colorbar on the right of the lower panels. {\it Lower
      panels:} projected galaxy distribution around reference high-z
    QSOs. The depth of the projection is 20 cMpc. Purple pluses and
    crosses mark the position of AGN companions with L$_{\rm bolo}$
    larger than 10$^{42.5}$ erg/s and 10$^{44}$ erg/s,
    respectively. In all panels, the grey shading correspond to the
    underlying projected distribution of dark matter, computed from
    the distribution of DMHs in the simulated
    volume.}\label{fig:env3d}
\end{figure*}

We extract a sample of high-z QSOs using the following selection. We
first select all QSOs at z>6 with bolometric luminosity above a
threshold of L$_{\rm bolo}$>10$^{46.25}$ erg/s. We then select among
those all model galaxies with Black Hole mass M$_{\rm BH}$ > 10$^{7.5}
\msun$ and stellar mass M$_{\star}$ > 10$^{9} \msun$, ending up with a
sample of 56 systems at 6<z<8. These cuts are meant to select the most
extreme objects hosting an accreting SMBH at high redshift. It is
worth stressing that this sample does not include all M$_{\star}$ >
10$^{9} \msun$ galaxies in this redshift range. Among these 56
sources, only 2 are satellite galaxies (their corresponding central
galaxies are not in the sample). For each high-z QSO we consider their
local environment, by selecting all galaxies closer than 10 cMpc
(comoving), at the redshift of detection. Finally, for each of our
selected high-z QSOs we reconstruct the redshift evolution of the main
progenitor and descendant along the corresponding merger trees, in
order to infer the various evolutionary paths and identify the low-z
counterparts.

In Fig.~\ref{fig:prop}, the blue histograms show the distribution of
the main properties of the QSO host galaxies at the redshift of
detection. Our sample covers a wide range of physical masses, both in
M$_{\star}$, M$_{\rm DM}$. In detail, \gaea~predicts host galaxies
with M$_{\star} \gtrsim$10$^{9.5} \msun$ living in parent DMHs with
M$_{\rm DM} \gtrsim$10$^{11.75} \msun$. These stellar and halo mass
values are smaller than expectation from hydro-dynamical simulations
(M$_{\star} \gtrsim$10$^{10} \msun$ -- e.g. \citealt{Marshall19} and
M$_{\rm DM} \gtrsim$10$^{12.3} \msun$ -- e.g. \citealt{Costa19}) but
consistent with results from other semi-analytic models
\citep{Fanidakis13}. The predicted M$_{\rm BH}$ distribution is
characterized by a steep slope, similar to the BH mass function of the
total population. This suggests that the presence of a bright QSO
depends more strongly on M$_{\rm BH}$ than on M$_\star$ or M$_{\rm
  DM}$.  Most of the objects are relatively star-forming and tend to
lie above the local M$_{\rm BH}$ - M$_{\star}$ relation. A critical
point in the \gaea~framework involves the different triggering
mechanisms for the AGNs, i.e. the relative role of disc instabilities
and mergers in sustaining the accretion onto the SMBHs. It is worth
noting that at such high redshifts, it is not always possible to
associate a unique triggering mechanism to each source in our
sample. Indeed, for 7 events in our sample (corresponding to 12.5 per
cent of the sources), we find that both channels can contribute to a
single AGN event. These complex events typically start with a minor
merger, that brings cold gas into the reservoir and fuels accretion
onto the central SMBH, as we described in Sec.~\ref{sec:simsam},
powering an AGN fainter than our selection limit. All the remaining
baryonic mass from the satellite is given to the disc of the remnant
galaxy. This additional material can destabilize the disc and trigger
an instability event, providing more cold gas to the reservoir around
the SMBH, eventually bringing the accretion above our selection
threshold. In the following, we include all events that start with a
merger under the ``merger'' label, even if most of the cold gas is
channeled in the reservoir as a consequence of the induced disc
instability. According to these definitions roughly half of the our
high-z QSO sample is triggered by disc instability events. The cyan
filled histogram refer to the distribution of the properties of host
galaxies for the subsample of QSO triggered by disc instability only
events. Overall, the comparison of the blue and cyan histograms
suggests that the most luminous events involving the more massive
SMBHs are mostly connected to galaxy mergers. We stress that all
merger events connected with the sources in our sample can be
classified as minor mergers (i.e. baryonic mass ratio between the
merging galaxies below 1/3).


\section{Environment of high-z QSOs in GAEA}
\label{sec:env}
\begin{figure}
  \centerline{ \includegraphics[width=9cm]{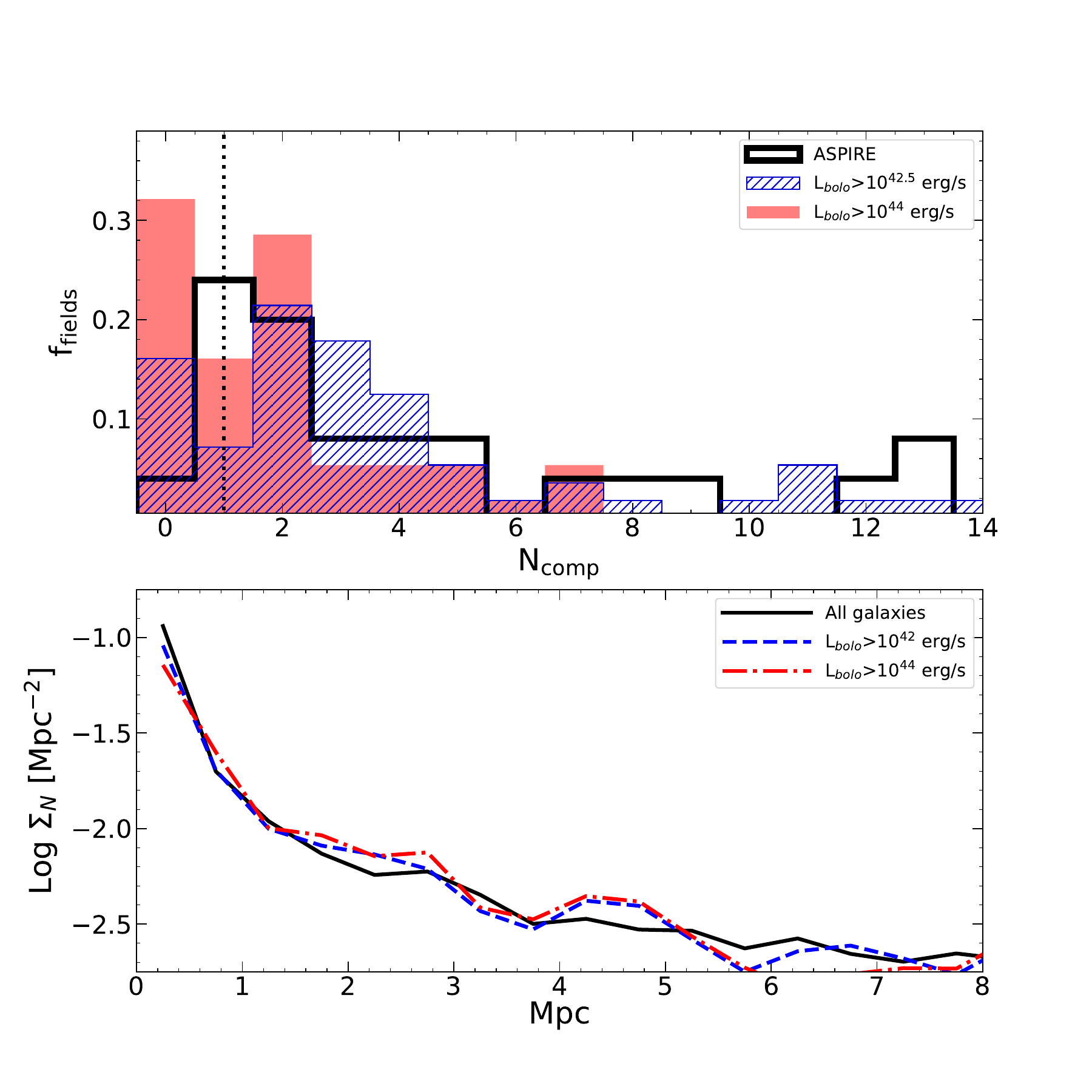}}
  \caption{ {\it Upper panel:} Fraction of fields $f_{\rm fields}$
    hosting a given number of M$_\star$>10$^{8} \msun$ AGN companions
    brighter than L$_{\rm bolo}$> 10$^{42.5}$ (blue dot-dashed
    histogram) and 10$^{44}$ erg/s (red dashed histogram and Orange
    filled area). The black histogram shows the fraction of fields in
    the ASPIRE survey, hosting a given number of [O$_{\rm III}$]
    emitters \citep{Wang26}. {\it Lower panel:} M$_\star$>10$^{8}
    \msun$ galaxy surface number density distribution as a function of
    projected galaxy separation from the central QSO. Solid black, red
    dashed and blue dot-dashed lines refer to the total galaxy
    population and the AGN companions (defined using a with L$_{\rm
      bolo}$> 10$^{42.5}$ and 10$^{44}$ erg/s AGN luminosity
    threshold) respectively.}\label{fig:distro}
\end{figure}

In order to characterize the environment of high-z QSOs we start from
the average number of galaxies (N$_{\rm comp}^{\rm rand}$) within a
spherical volume of 10 cMpc radius, defined from the position of
$\sim$20000 test galaxies randomly selected in the \pms~volume. For
the sake of these estimates only M$_\star$>10$^{8} \msun$ galaxies
have been considered as both targets and companions, and the procedure
has been repeated for all considered redshifts. We then compare these
values with the corresponding number of companions within the same
volume around bright QSOs in our sample (N$_{\rm comp}^{\rm QSO}$) and
we define a relative overdensity $\delta_{10}$ as:

  \begin{equation}
    \delta_{10} = N_{\rm comp}^{\rm QSO} / N_{\rm comp}^{\rm rand}
  \end{equation}

We thus use $\delta_{10}$ to rank the bright QSO environments. We find
that {\gaea} high-z bright QSOs live in regions characterized by a
wide $\delta_{10}$ interval, ranging from crowded regions with a large
number of companions (fourth quartile of the distribution,
corresponding to $\delta_{10} \sim$10) to regions comparable to the
mean random density field (first quartile, $\delta_{10} \sim$1).

We then consider two representative sources from the first and fourth
quartile of $\delta_{10}$. In detail, these two sources have been
selected at the same redshift (z$\sim$7.64) and roughly at the same
L$_{\rm bolo} \sim$ 10$^{46.6}$ erg/s. We highlight the properties of
these two objects in each panel of Fig.~\ref{fig:prop}, by means of
vertical dashed and dotted lines. The two selected object also have a
similar M$_{\star}$ ($\sim$0.06 dex difference) and M$_{\rm DM}$
($\sim$0.05 dex difference), while their M$_{\rm BH}$ differ by
$\sim$0.25. In Fig.~\ref{fig:env3d} we show the distribution of
companion galaxies with M$_{\star}$ > 10$^{8} \msun$, and the
underlying DM density distribution, in cubic boxes of 20 cMpc per side
centred on the reference QSO. In the upper panels we show the
tridimensional distribution around the selected high-z QSO, while in
the lower panels we show an arbitrary 2d projection (the depth of the
projection still correspond to 20 cMpc). In each panel the size of the
points is proportional to the stellar mass of the model galaxy, and
each of them is colour-coded according to its SFR, as indicated in the
the colorbar on the right of each panel. The QSO host on the left
(QSO-A hereafter) lives in a significant overdensity ($\delta_{10}
\sim$8 belonging to the fourth quartile of the distribution) that will
evolve into a massive M$_{\rm DM} \sim$ 10$^{14.9} \msun$ cluster at
z$\sim$0 (see Sec.~\ref{sec:z0} for more details), while the host
galaxy on the right (QSO-B hereafter) lives in a slightly underdense
region ($\delta_{10} \sim$0.6 belonging to the first quartile of the
distribution) and its activity has been triggered by a disc
instability event in a gas rich galaxy disc. In the lower panels, we
show the corresponding projected distribution of galaxies on x-y
plane. In these plots we also mark the position of active companions
with crosses and pluses (defined as those sources with L$_{\rm bolo}$
larger than 10$^{42.5}$ erg/s and 10$^{44}$ erg/s, respectively). The
abundance of active companions has been already used to search for
potential proto-cluster fields \citep[see e.g.][]{Gilli03}.
\begin{figure*}
  \centerline{ \includegraphics[width=18cm]{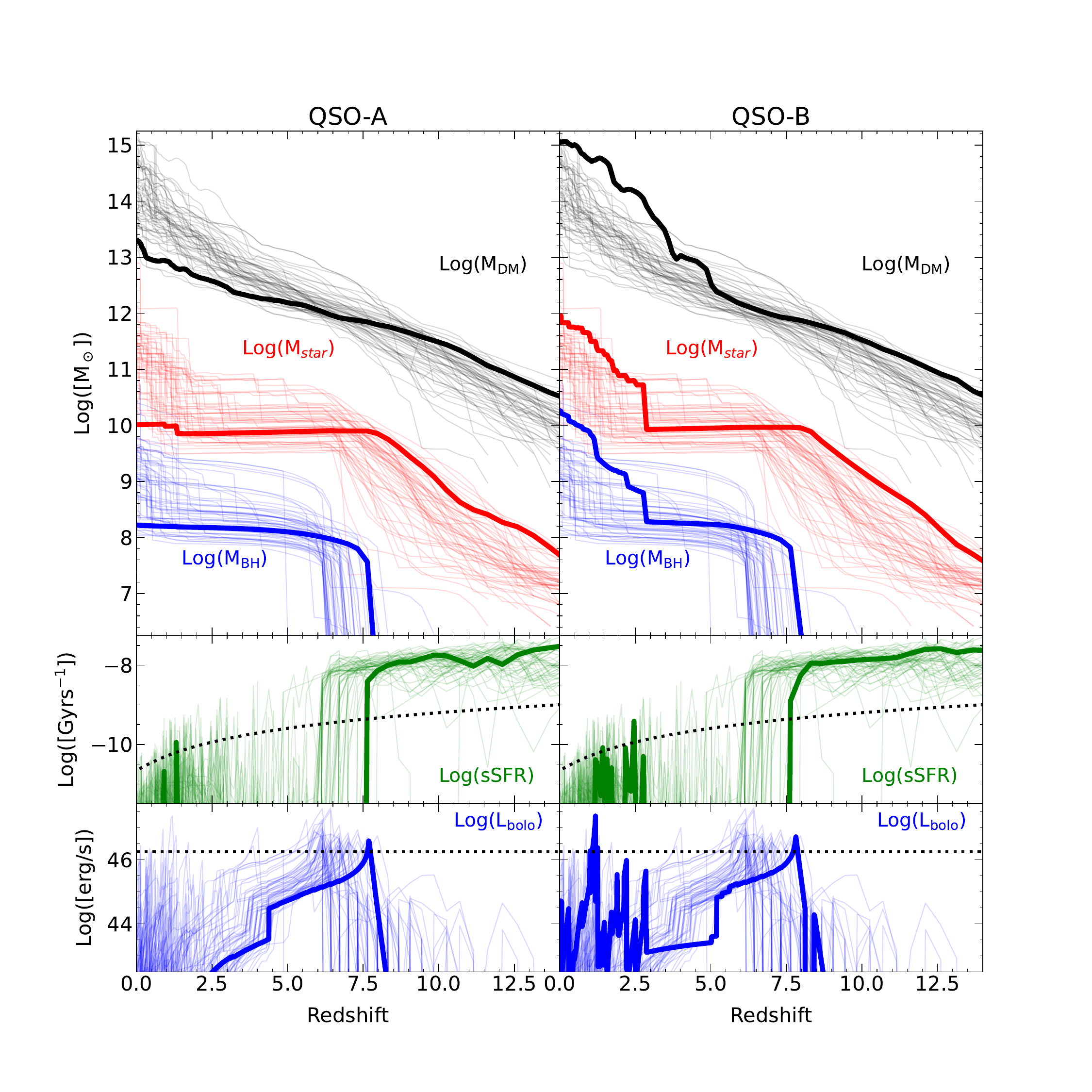}}
  \caption{Redshift evolution of selected properties of the host
    galaxies for the two reference high-z QSOs (thick lines). In all
    panels, the evolutionary tracks for the entire sample of high-z
    QSOs is shown with thin lines. Left Panels correspond to the
    redshift evolution for QSO-A, while the right panel for QSO-B (see
    text for more details). Upper panels show the redshift evolution
    of the parent DMH mass, stellar mass and SMBH mass. Lower panels
    show the redshift evolution for the sSFR and bolometric
    luminosity. In the sSFR panel, the dotted line represent a
    classical threshold for defining star-forming galaxies in
    theoretical models. In the L$_{\rm bolo}$ panel, the horizontal
    line marks the detection threshold for high-z QSOs we assume in
    this work.}\label{fig:evo}
\end{figure*}

For a more quantitative analysis on the distribution of active
companions, we compare \gaea~predictions with the results from the
ASPIRE program \citep{Wang26}, a galaxy redshift survey targeting
fields around 25 z$\gtrsim$6.5 QSOs with JWST and ALMA. Within ASPIRE,
O$_{\rm III}$ emitters have been detected both along the line of sight
and close to the reference source. It is important to keep in mind
that this O$_{\rm III}$ line survey selects a mixed population
including both star-forming galaxies and AGNs and additional emission
lines would be necessary to disentangle sources dominated by emission
from the central BH. On the other hand our predictions refer only to
the luminosity connected of AGN activity, neglecting the contribution
of star formation. In general, one may expect the brightest emitters
to be AGN dominated, but at the assumed selection threshold for the
survey (i.e. L$_{\rm bolo}$>10$^{42}$ erg/s), an important
contribution of star-forming galaxies has to be expected.

With these caveats in mind, we show in the upper panel of
Fig.~\ref{fig:distro} the distribution of the 56 mock fields in terms
of number of active companions (at different L$_{\rm bolo}$
threshold). The resolution limit of \gaea~predictions based on the
\pms~matches well the expected stellar mass detection limit for
star-forming galaxies in ASPIRE (i.e. M$_\star \gtrsim$10$^{8} \msun$
\citealt{Champagne25b}). In the \gaea~realizations, there are
typically less than 3 companions brighter than 10$^{44}$ erg/s and no
more than 7, while when considering a lower luminosity cut
(10$^{42.5}$ erg/s), the number of companions varies from zero to
15. The model predicts that almost half of the fields centred around
high-z QSOs should contain at least 2 bright (L$_{\rm bolo}$>
10$^{44}$ erg/s) QSOs. The fraction of fields with more than 2 AGN
companion raises to $\sim$0.8 if the fainter L$_{\rm bolo}$ threshold
is considered. This result is in qualitative agreement with one of the
key findings of the ASPIRE program, i.e. the fact that the number of
companion O$_{\rm III}$ emitters varies significantly from field to
field (black empty histogram in the upper panel of
Fig.~\ref{fig:distro}). In detail, 13 surveyed fields (corresponding
to a 0.52 fraction) include more than 2 O$_{\rm III}$ emitters, 2
ASPIRE fields include 13 sources, and a single field has 20
emitters. The clustering properties of O$_{\rm III}$ emitters in the
ASPIRE fields suggest a characteristic halo mass of M$_{\rm DM}
\sim$10$^{10.55} \msun$ \citep{Huang26}, in agreement with the median
value of the host halo mass of active companions in \gaea~mock fields
(10$^{11.11} \msun$). The same clustering analysis provides an
estimate for the characteristic halo mass of the reference bright QSOs
of M$_{\rm DM} \sim$10$^{12.2} \msun$ \citep{Wang26, Huang26}, also in
agreement with the distribution shown in Fig.~\ref{fig:prop}.

We also consider the mean projected number density of
M$_\star$>10$^{8} \msun$ companions galaxies in our mock fields
(Fig.~\ref{fig:distro} - lower panel). Black solid, blue dashed and
red dot-dashed lines show the resulting distributions as a function of
distance from the reference QSO for the all companions (both active
and inactive) and for AGN companions above the usual luminosity
thresholds. All distributions have been normalized to the total number
of sources in each sample to ease the comparison. The results
highlight that AGN companions do not show any clear signal for a
different distribution (i.e. a stronger clustering) with respect to
the overall galaxy population around these high-z QSOs.

\section{Redshift evolution}\label{sec:evo}
\begin{figure}
  \centerline{ \includegraphics[width=9cm]{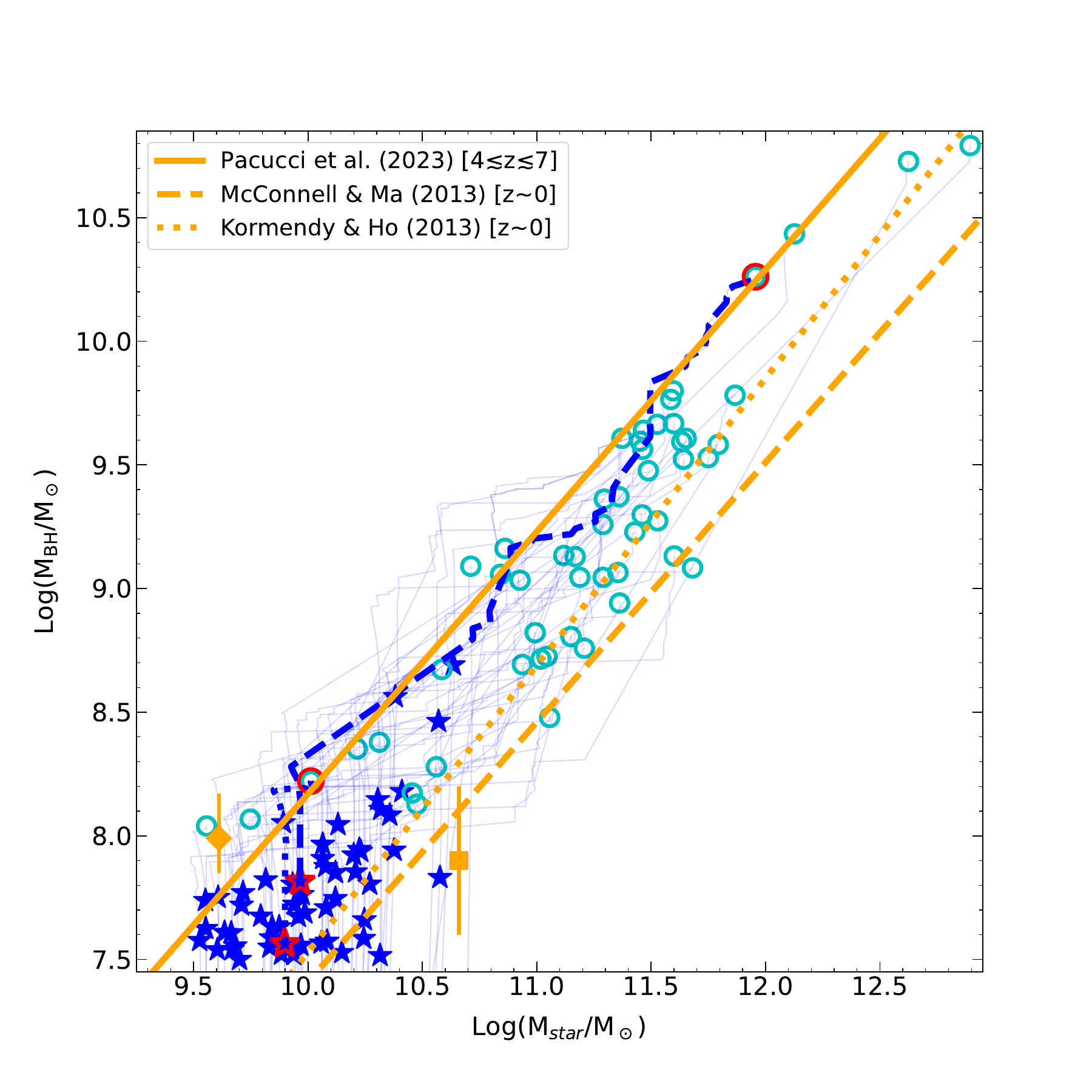}}
  \caption{Evolution of the M$_{\rm BH}$-M$_{\star}$ relation for the
    host galaxies of high-z QSOs. Blue stars represent sources in the
    {\gaea} z>6 QSO sample. Thin lines show the evolution of the
    relation following the direct descendants of the host galaxy. Cyan
    cycle mark the position of their corresponding z=0
    descendants. Red open symbols and blue thick lines show the
    evolution for QSOA-A and QSO-B (dotted and dashed line
    respectively). Orange dots show z>6 sources in the
    \citet{Harikane23} and \citet{Maiolino24} samples (diamond and
    square respectively). Orange lines refer to observed relations at
    different redshifts, as estimated by \citet[][at
      4<z<7]{Pacucci23}, \citet[][at z$\sim$0]{KormendyHo13} and
    \citet[][at z$\sim$0]{McConnellMa13}.}\label{fig:mago}
\end{figure}

We then consider the evolutionary history of the high-z QSO host
galaxies as predicted by \gaea. In Fig.~\ref{fig:evo}, we collect all
evolutionary tracks for the 56 objects in our sample (light lines). In
detail, we consider the redshift evolution of mass components (M$_{\rm
  DM}$, M$_{\star}$, M$_{\rm BH}$), as well as the evolution of the
specific SFR (sSFR = SFR/M$_\star$) and L$_{\rm bolo}$. Overall, these
evolutionary tracks span a wide range of possible histories, both in
terms of mass assembly and activity. In Fig.~\ref{fig:evo} we also
highlight with bold lines the tracks corresponding to the same two
representative sources selected in Fig.~\ref{fig:env3d}. As discussed
in previous section, the two reference QSOs have similar values for
L$_{\rm bolo}$, M$_{\star}$, M$_{\rm DM}$ and M$_{\rm BH}$ at the
detection redshift (z$\sim$7.64). Nonetheless, their evolution at
lower redshift is strikingly different. In the right-hand panels, we
show that the evolution for QSO-B after the main triggering event (a
disc instability) is rather quiet, the host galaxy rapidly becomes
quiescent and it does not form a significant fraction of its stars at
later times. Despite being one of the most massive systems in the
\pms~at z$\sim$7.64 with M$_{\star} \sim$10$^{10} \msun$, the host
galaxy does not grow significantly in mass down to z=0.

In the left-hand panels of Fig.~\ref{fig:evo} we show QSO-A, living in
a dense environment at z$\sim$7.64, and characterized by a more
complex evolutionary history. Also for this source the main triggering
event is a disc instability. The host galaxy SFR rapidly drops below
the assumed threshold for star forming galaxies (similarly to QSO-B)
and it does not form stars at a relevant\footnote{We define a model
galaxy as star forming if its specific SFR is larger than 0.3/t$_H$,
where t$_H$ is the Hubble time at the redshift under consideration
\citep{Franx08}. This limit is marked by a dotted line in
Fig.~\ref{fig:evo}.} rate down to z$\sim$2.5 (thus being likely
classified as a massive red and dead galaxy in this redshift
interval). Then it suffers a series of merger events, leading to the
formation of a Bright Central Galaxy (BCG) of a M$_{\rm
  DM}\sim$10$^{15} \msun$ galaxy cluster. This merger series is also
associated to a very bright QSO phase at z$\sim$1. It is worth noting
that during this merger phase the high-z QSO host galaxy does not
correspond to the most massive galaxy in its environment (i.e. the
main progenitor of the BCG). We stress that the two examples shown in
Fig.~\ref{fig:evo} are representative for the first and fourth
quartiles of $\delta_{10}$.  Overall, the ensemble of light lines in
Fig.~\ref{fig:evo} shows the intrinsic stochasticity of mass assembly
in a hierarchical Universe driven by mergers, as already widely
discussed in the literature \citep[see e.g.][]{DeLuciaBlaizot07,
  Trenti08, Overzier09, Angulo12, DeLucia25}.

We also consider the position of our model galaxies in the M$_{\rm
  BH}$-M$_\star$ plane in Fig.~\ref{fig:mago}. Recent estimates based
on JWST observations at 4<z<7 \citep{Pacucci23} suggest an evolution
of the relation with respect to local estimates \citep{KormendyHo13,
  McConnellMa13}, that bears relevant consequences for our
understanding of the co-evolution of SMBH and their host galaxies.  In
particular, high-z observations show that SMBHs at these redshifts are
systematically more massive at fixed stellar mass than the expectation
from the local relation \citep{Harikane23, Maiolino24}. Most of our
bright high-z QSOs (blue stars) lie between the local and the high-z
relations, thus showing that also in {\gaea} SMBHs tend to grow faster
than their host galaxies in the early Universe. It is important to
stress that our theoretical sources and observational samples used to
estimate the high-z M$_{\rm BH}$-M$_\star$ do not match well in
stellar mass. Indeed, most of the observational estimates correspond
to galaxies with lower stellar masses (i.e. M$_\star$ $\lesssim$ 10$^9
\msun$) with respect to the predicted hosts for bright QSOs. In
Fig.~\ref{fig:mago} we mark the location of the only two z>5 objects
in the \citet{Harikane23} and \citet{Maiolino24} samples with
comparable M$_\star$. Therefore, it remains unclear if the lack of
extreme outliers in the relation (like those reported in
\citealt{Juodzbalis24} and \citealt{Bogdan24} at lower stellar mass)
is indeed a problem.

In Fig.~\ref{fig:mago}, we also show with thin cyan lines the
evolution of each high-z QSO in the M$_{\rm BH}$-M$_\star$ plane. The
empty cyan circles represent the position of each descendant at
z=0. At z>4, all selected model QSOs still show a faster growth in
M$_{\rm BH}$ than in M$_\star$, which keeps them above both the high-z
and the local relation. Objects like QSO-A (blue dotted line) do not
evolve any further at lower redshifts, while sources like QSO-B (blue
dashed line) tend either to evolve along the \citet{Pacucci23}
relation (if they have lower-z reactivation events - please note that
the solid line in Fig.~\ref{fig:mago} represents an extrapolation from
the mass range this relation has been originally defined), or move
back to the local relation (if their evolution is dominated by galaxy
mergers with no additional bright QSO events -
\citealt{JahnkeMaccio11}). At z$\sim0$ the descendants of the high-z
QSOs (cyan circles) represent a galaxy population whose SMBH tend to
lie above the local M$_{\rm BH}$-M$_\star$ relation, although a
significant fraction ($\sim$60 per cent) is within the quoted 0.3 dex
around the observed relation.

\section{The relation between high-z QSO and Large Scale Structure}
\label{sec:z0}
\begin{figure}
  \centerline{ \includegraphics[width=8cm,height=12cm]{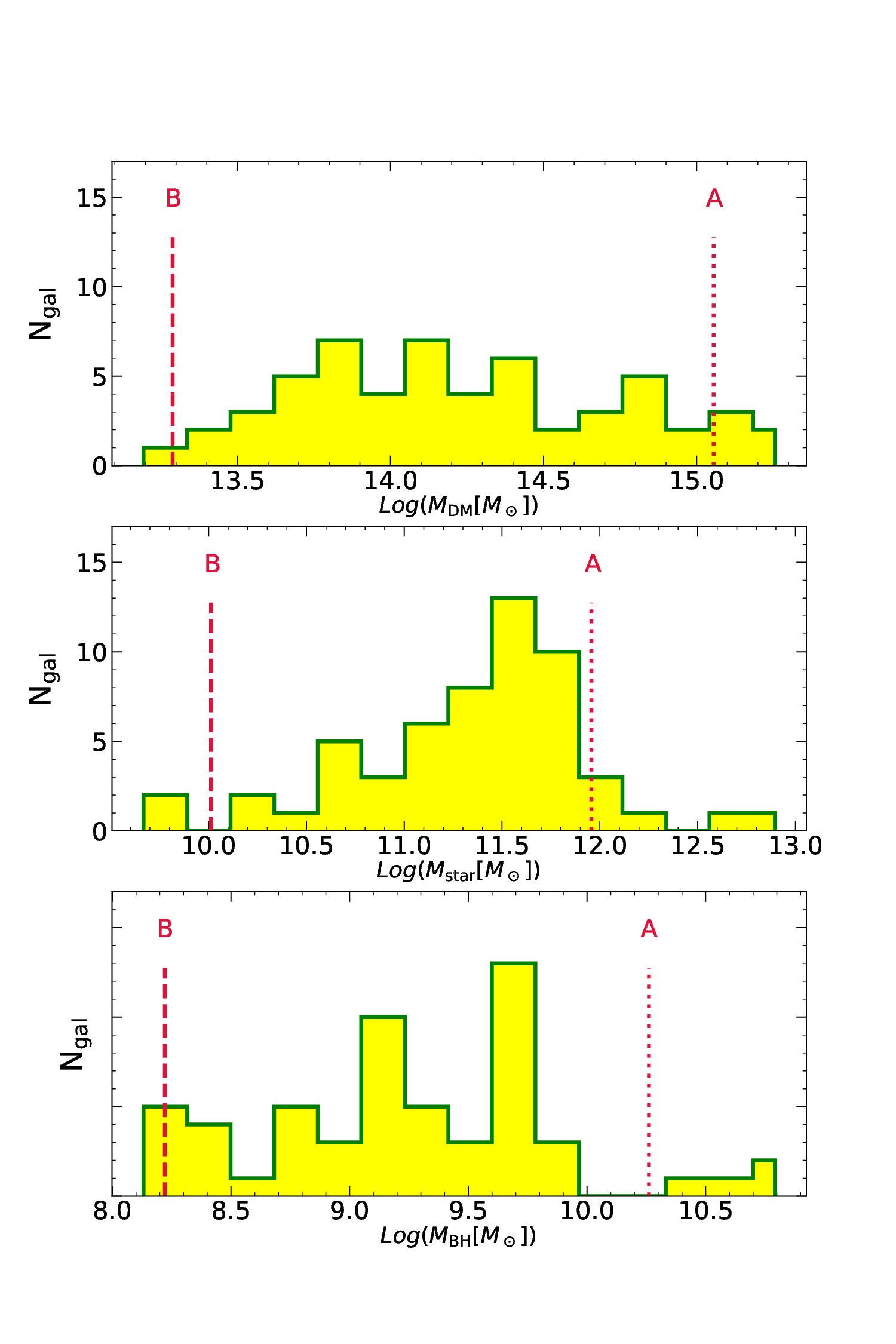}}
  \caption{Properties for main descendants of high-z QSO
    hosts. Vertical dashed and dotted lines mark the properties of
    QSO-A and QSO-B descendants.}\label{fig:finprop}
\end{figure}

In order to better quantify the relationship between high-z QSOs and
their evolution into the galaxy population at z$\sim$0, we show the
distribution of some selected properties of their descendant galaxies
at z$\sim$0 in Fig.~\ref{fig:finprop}. When considering the properties
of the descendant population, the final distributions for the mass
components (M$_{\rm DM}$, M$_{\star}$, M$_{\rm BH}$) are still quite
broad. As already discussed in our previous analysis, QSO-A and QSO-B
lie at the extremes of the z$\sim$0 distributions. Our main
conclusions do not change if we restrict the sample of model galaxies
only to the hyper-luminous QSOs (i.e.  L$_{\rm bolo}$>10$^{47}$ erg/s,
see also \citealt{Salvestrini25}). Only 32 out of 56 objects in our
sample (57 per cent) end up in galaxy clusters (M$_{\rm DM}$>10$^{14}
\msun$) at low redshift, while the remaining fraction lives in massive
groups (M$_{\rm DM}$>10$^{13} \msun$). Moreover, our sample does not
trace all the more massive structures in the simulation box. Indeed,
we identify 45 galaxy clusters with M$_{\rm DM}$>10$^{15} \msun$ in
the \pms~at z=0. Only 4 of these include descendants of high-z QSOs
(such as QSO-A), which corresponds to only 9 per cent of the z=0
galaxy clusters\footnote{This figure decreases with M$_{\rm DM}$,
i.e. only 0.6 per cent of galaxy clusters with M$_{\rm DM}$>10$^{14}
\msun$ and 2 per cent of galaxy clusters with M$_{\rm DM}$>10$^{14.5}
\msun$ in the \pms~include a descendant of a high-z QSO from our
sample.} None of the high-z QSO in our sample is the main progenitor
of a central BCG, but rather belongs to another structure merging with
the main progenitor of the cluster halo at later times. Therefore,
even if not the more massive galaxy along the evolutionary path of the
BCG, the high-z QSO usually belongs to the more active evolutionary
branch. However, it is quite telling, that the largest fraction (more
than 90 percent) of z$\sim$0 M$_{\rm DM}$>10$^{15} \msun$ clusters in
the \pms, do not include any galaxy corresponding to the descendant of
our bright high-z QSOs. We study the evolution of the properties of
the BCG at different epochs and we find that the main reason for this
is that their more active phase typically peaks at lower redshift (in
a large z$\sim$3-5 range) than our selection at z>6. Moreover, 8
high-z QSO in our sample (14 per cent) correspond to a satellite
descendant at z=0, and 2 of them (3.5 per cent) to an orphan galaxy
(i.e. a model satellite galaxy whose substructure has been stripped
below the resolution of the merger tree). These numbers are consistent
with the results of \citet{Fanidakis13}, based on the {\sc galform}
model.

\begin{figure}
  \centerline{ \includegraphics[width=9cm]{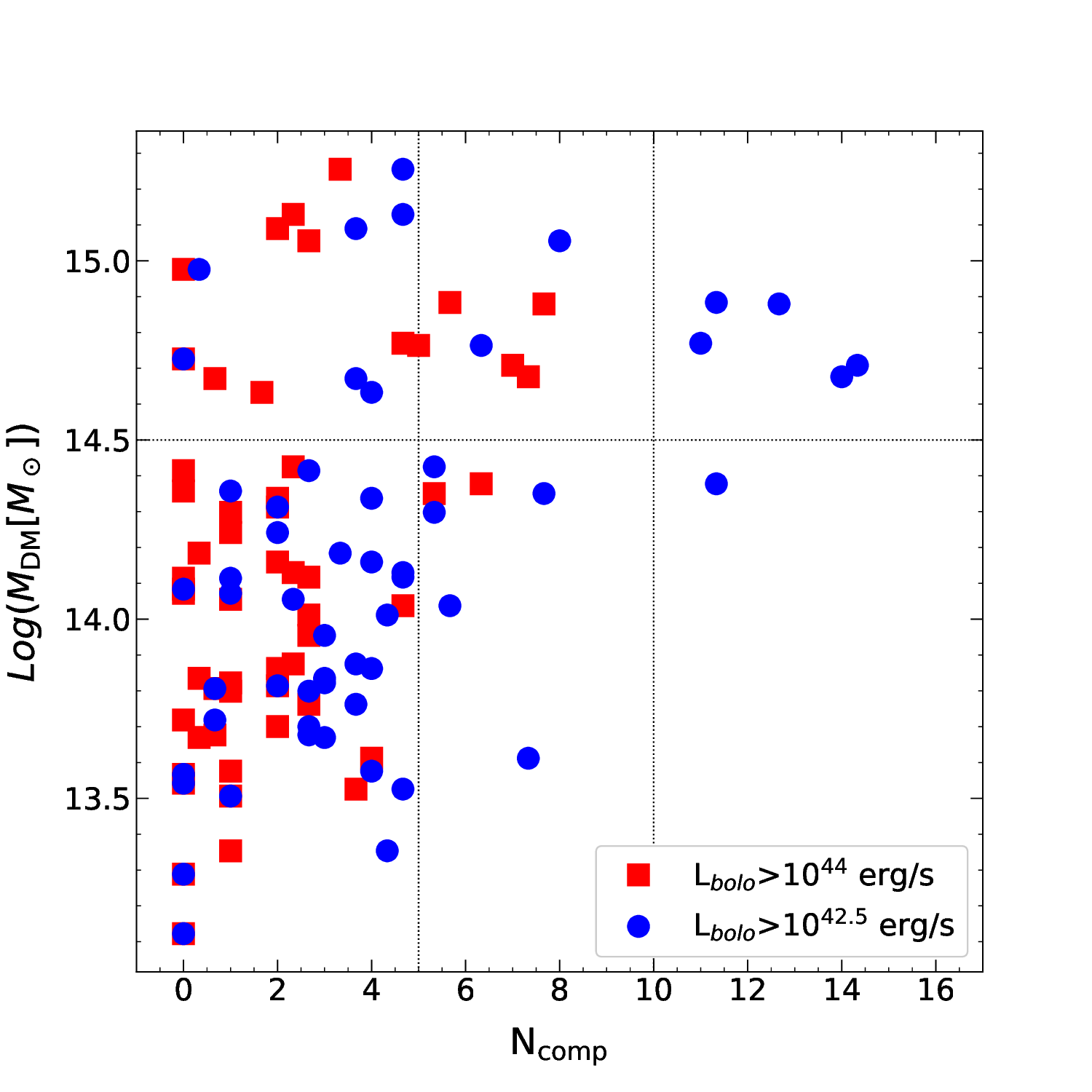}}
  \caption{Parent DMH mass M$_{\rm DM}$ of z=0 descendant galaxies of
    high-z bright QSOs as a function of AGN companions (defined over a
    10 cMpc scale). Blue circles and red square refer to AGN
    companions brighter than L$_{\rm bolo}$> 10$^{42.5}$ erg/s and
    10$^{44}$ erg/s, respectively.}\label{fig:evocol}
\end{figure}

In Fig.~\ref{fig:evocol} we relate the z$\sim$0 parent DMH mass of the
descendants of high-z bright QSOs with the number of their AGN
companions. As in Sec.~\ref{sec:env}, we consider two different
luminosity thresholds for defining the active galaxy companion
population, defined in cubes of size 20 cMpc centred on the reference
QSO. Results highlight that z>6 QSOs with bright (L$_{\rm bolo}$>
10$^{44}$ erg/s) AGN companions of the order of $\sim5$ can be
considered as candidates for proto-cluster environments (defined as
regions ending up in a M$_{\rm DM}$>10$^{14.5} \msun$
structure). These numbers should be increased to $\sim$10, if the
fainter AGN detection threshold is considered. Nonetheless, the
relation between M$_{\rm DM}$ and the number of active companions
shows a relevant scatter on both axis: proto-cluster regions around
high-z QSOs may also have a relatively low number of active
companions.

\section{Summary}
\label{sec:discconcl}

In this paper, we present predictions from the latest version of the
\gaea~model, run on merger trees extracted from the
\pmill~simulation. We focus our analysis on a sample of mock z>6 QSOs,
defined as sources with L$_{\rm bolo}$>10$^{46.25}$ erg/s; M$_{\rm
  BH}$ > 10$^{7.5} \msun$ and M$_{\star}$ > 10$^{9} \msun$. These
selection cuts are meant to identify the most extreme objects in terms
of SMBH accretion in our simulated box. We consider both the
environment (defined as the number of companion galaxies within a
cubic box of 20 cMpc size) around the selected sources, as
well as their lower redshift evolution. Our main results can be
summarized as follows:

\begin{itemize}
\item[$\ast$]{Bright high-z QSOs in \gaea~are triggered in a variety
  of environments, ranging from galaxy overdensities (defined on a
  scale of ~10 cMpc around the reference source) that may collapse in
  galaxy clusters at $z\sim$0, to regions of more average density
  evolving in a group-like environment (i.e. a few 10$^{13} \msun$) by
  z$\sim$0. Roughly half of the sources in our sample have been
  triggered by disc instability events. The brightest events in more
  massive galaxies and DM haloes originate from galaxy mergers.}
\item[$\ast$]{Roughly half of the mock fields contain AGN companions
  with bolometric luminosity in excess of 10$^{44}$ erg/s and stellar
  mass above our nominal resolution (10$^8 \msun$). These figures
  increase if a fainter luminosity threshold is adopted to select AGN
  companions (e.g. 80 percent for L$_{\rm bolo}$>10$^{42.5}$ erg/s).}
\item[$\ast$]{The evolution of the host galaxies of bright high-z QSOs
  show a high degree of variability and descendants are predicted to
  span a wide range of physical properties. All host galaxies stop
  forming stars as the result of the QSO activity. Some of the
  selected sources do not experience other major star forming events
  after the main triggering event, thus ending up in a ``red and
  dead'' galaxy at lower redshifts. However, $\sim$40 per cent of
  model sources are characterized by a rejuvenation phase, with
  subsequent star formation and AGN events at later times.}
\item[$\ast$]{Local descendants of high-z QSOs live in a wide range of
  environments. Although they all live in M$_{\rm DM}$ > 10$^{13}
  \msun$, only a small fraction (4 out of 56, 7 percent) ends up in
  M$_{\rm DM}$ > 10$^{15} \msun$ clusters of the box. Conversely, only
  a small fraction (4 out of 45, 9 percent) of z$\sim$0 M$_{\rm DM}$ >
  10$^{15} \msun$ clusters in the \pms, have a bright high-z QSO
  within their progenitors. Our results suggests that observed bright
  high-z QSOs are poor tracers of the progenitors of local massive
  clusters, and additional information (such as the number of active
  galaxy companions) is needed to select the most promising
  candidates.}
\end{itemize}

These results are in qualitative agreement with recent finding from
the ASPIRE program, which highlight a relevant field to field
variation in the number of AGN companions around z$\gtrsim$6.5 bright
QSOs, although our results suggest the need for follow-up observations
covering a large field around target QSOs. \gaea~predictions do not
support the assumption of a one-to-one connection between the
brightest (L$_{\rm bolo}$>10$^{46.25} erg/s$) QSO found at z>6 and the
progenitors for the most massive M$_{\rm DM}$ > 10$^{14.5} \msun$
clusters at z$\sim$0 (see also \citealt{Overzier09} and
\citealt{Eilers24}). The evolution of their host galaxies shows a
variety of paths, which can be broadly tied to their high-z
environment, defined as the number of active companions in our mock
fields (defined on a $\sim$10 cMpc scale). Despite a tendency for
high-z QSOs with a large number of active companions to end-up in
massive structures at z$\sim$0, we find that their host galaxies do
not always correspond to the main progenitor of the more massive
z$\sim$0 galaxies in these large haloes in the \pms.

\begin{acknowledgements}
  We thank Feige Wang for sharing results from the ASPIRE program. An
  introduction to {\gaea}, a list of our recent work, as well as
  datafile containing published model predictions, can be found at
  \url{https://sites.google.com/inaf.it/gaea/home}. We acknowledge the
  use of INAF-OATs computational resources within the framework of the
  CHIPP project \citep{Taffoni20} and the INAF PLEIADI program
  (\url{http://www.pleiadi.inaf.it}). FF, RD, GDL and OC acknowledge
  the support and hospitality of the Institute for Fundamental Physics
  of the Universe – IFPU, that has been instrumental for to the
  development of this work, under the Team Research Program ``The
  environment of Quasars at the Cosmic Dawn'' held on 17-21 February
  2025. FF acknowledges support from the Next Generation European
  Union PRIN 2022 ``20225E4SY5 - From ProtoClusters to Clusters in one
  Gyr''. RD acknowledges support from the INAF GO 2022 grant ``The
  birth of the giants: JWST sheds light on the build-up of quasars at
  cosmic dawn'', and by the PRIN MUR ``2022935STW'', RFF M4.C2.1.1,
  CUP J53D23001570006 and C53D23000950006.
\end{acknowledgements}

\bibliographystyle{aa} 
\bibliography{fontanot} 

\end{document}